\documentclass[%
 reprint,
superscriptaddress,
 amsmath,amssymb,
 aps,
prm,
]{revtex4-1}

\usepackage{graphicx}
\usepackage{epstopdf}
\usepackage{dcolumn}
\usepackage{bm}
\usepackage{gensymb}
\usepackage{upgreek}
\usepackage{hyperref}
\usepackage{tabularx}
\usepackage{xcolor}

\begin{document}
\author{Brandon Wilfong}
\affiliation{Physics Department, United States Naval Academy, Annapolis, MD 20899, USA}
\author{Vaibhav Sharma}
\affiliation{Mechanical and Nuclear Engineering, Virginia Commonwealth University, Richmond, VA 23220, USA}
\author{Omar Bishop}
\affiliation{Mechanical and Nuclear Engineering, Virginia Commonwealth University, Richmond, VA 23220, USA}
\author{Adrian Fedorko}
\affiliation{Physics Department, Northeastern University, Boston, MA 02115, USA}
\author{Don Heiman}
\affiliation{Physics Department, Northeastern University, Boston, MA 02115, USA}
\affiliation{Plasma Science and Fusion Center, MIT, Cambridge, MA, 02139, USA}
\author{Radhika Barua}
\affiliation{Mechanical and Nuclear Engineering, Virginia Commonwealth University, Richmond, VA 23220, USA}
\author{Michelle E. Jamer}
 \email{jamer@usna.edu}
\affiliation{Physics Department, United States Naval Academy, Annapolis, MD 20899, USA}

\date{\today}

\begin{abstract}

Fe$_{3}$Ga$_{4}$ displays a complex magnetic phase diagram that is sensitive and tunable with both electronic and crystallographic  structure changes. In order to explore this tunability, vanadium-doped (Fe$_{1-x}$V$_{x}$)$_{3}$Ga$_{4}$ has been synthesized and characterized. High-resolution synchrotron X-ray diffraction and Rietveld refinement show that  samples up to 20\% V-doping remain isostructural to Fe$_{3}$Ga$_{4}$ and display a linear increase in unit cell volume  as doping is increased. Magnetic measurements reveal a suppression of the antiferromagnetic helical spin-density wave (SDW) with V-doping, revealed by changes in both the low-temperature ferromagnetic-antiferromagentic (FM-AFM) transition (T$_{1}$) and high-temperature AFM-FM transition (T$_{2}$). At 7.5\% V-doping, the metamagnetic behavior of the helical AFM SDW phase is no longer observed. These results offer an avenue to effective tuning of the magnetic order in Fe$_{3}$Ga$_{4}$ for devices, as well as increased understanding of the magnetism in this system.

\end{abstract}

\title{The effect of vanadium substitution on the structural and magnetic properties of (Fe$_{1-x}$V$_{x}$)$_{3}$Ga$_{4}$}

\maketitle


Materials with metallic antiferromagnetic behavior at room temperature have been a focus of recent research due to the unique interplay of electron spin and charge.\cite{Siddiqui, zhang2014, gomonay} These materials are especially important in AFM spintronic devices if they are metallic.\cite{Jungwirth, Hoffmann, Baltz} A material of interest is intermetallic Fe$_{3}$Ga$_{4}$, which displays a complex magnetic phase evolution with respect to temperature and magnetic field.\cite{Wilfong2021, Mendez, Samatham, Kawamiya} The ground state of Fe$_{3}$Ga$_{4}$ is ferromagnetic (FM), which transitions to an intermediate AFM phase at T$_{1}$~$\sim$~68~K that exists over a wide temperature range, and transitions to a high-temperature FM phase at T$_{2}$~$\sim$~360~K. The intermediate AFM phase exists at room-temperature, is metallic, and has been explored more in-depth in recent work, determining that the AFM order is a helical spin-density wave (SDW).\cite{Wilfong2021, Afshar, WilfongSC, Wu} In addition to metallic behavior, the helical AFM/SDW exhibits complex metamagnetic evolution that is integral to its exotic properties such as the topological Hall effect.\cite{Afshar, Mendez} 

For potential applications, the magnetic phase transition from an AFM state to an FM state near room temperature ($\sim$~360~K) is of key interest, such as in FeRh.\cite{Loving, Barua, Bennett2016, Cress} It is important to understand how to tune the magnetic phase transition temperatures by external parameters. Additionally, from a fundamental point-of-view, probing external parameters' effect on these magnetic transitions will be useful to understand how and why the magnetic phases in this material manifest. \cite{Moriya, Moriya2, Feng} Previous work has shown that external pressure is an effective means to tune the magnetic transition temperatures;\cite{DujinPressure, Wilfong2021, DuijnThesis, WuThesis} however, this has not been systematically linked to structural and/or electronic changes induced by pressure. Previous work determined the effect of doping of other metals on the magnetic phase diagram of Fe$_{3}$Ga$_{4}$ through Fe site and Ga site substitution.\cite{Al1995, Al2000, Duijn, Kobeissi, Al1998} From these studies, it is clear that the  helical SDW is sensitive to effects caused by doping similar to electronic and crystallographic structural changes, and the incorporation of vanadium is studied here. 

In this work, V-doped (Fe$_{1-x}$V$_{x}$)$_{3}$Ga$_{4}$ was prepared with x = 0.025, 0.05, 0.075, 0.1, 0.15, 0.2, in order to explore how both structural and electronic changes affect the magnetic phases.  Previous work has shown that minute changes in crystal structure induced by annealing and pressure can change the magnetic transition temperatures, but details of V-doping are absent.\cite{Wilfong2021, Mendez}  Due to the electronic configuration of V, the addition of V as a dopant will lead to one of the largest possible expansions of the crystal lattice in the Fe$_3$Ga$_4$ compound, which could give insight on the effect of negative pressure in the lattice. We have performed high-resolution synchrotron X-ray diffraction analysis in order to track crystallographic changes concomitant with V-doping and probe the effect on magnetic properties. Additionally, previous work has shown the importance of both the electronic configuration and Fe-Fe magnetic interactions on the magnetic phase diagram of Fe$_{3}$Ga$_{4}$,\cite{Wilfong2021, Afshar, Mendez} and we seek to probe how V-doping on the Fe sites affects the magnetic properties. This current work offers another avenue for tuning the magnetic phases for potential applications as well as to better understand the mechanism for changes in magnetic order. 

\begin{figure*}[]
    \centering
    \includegraphics[width = 5in, height = 4in]{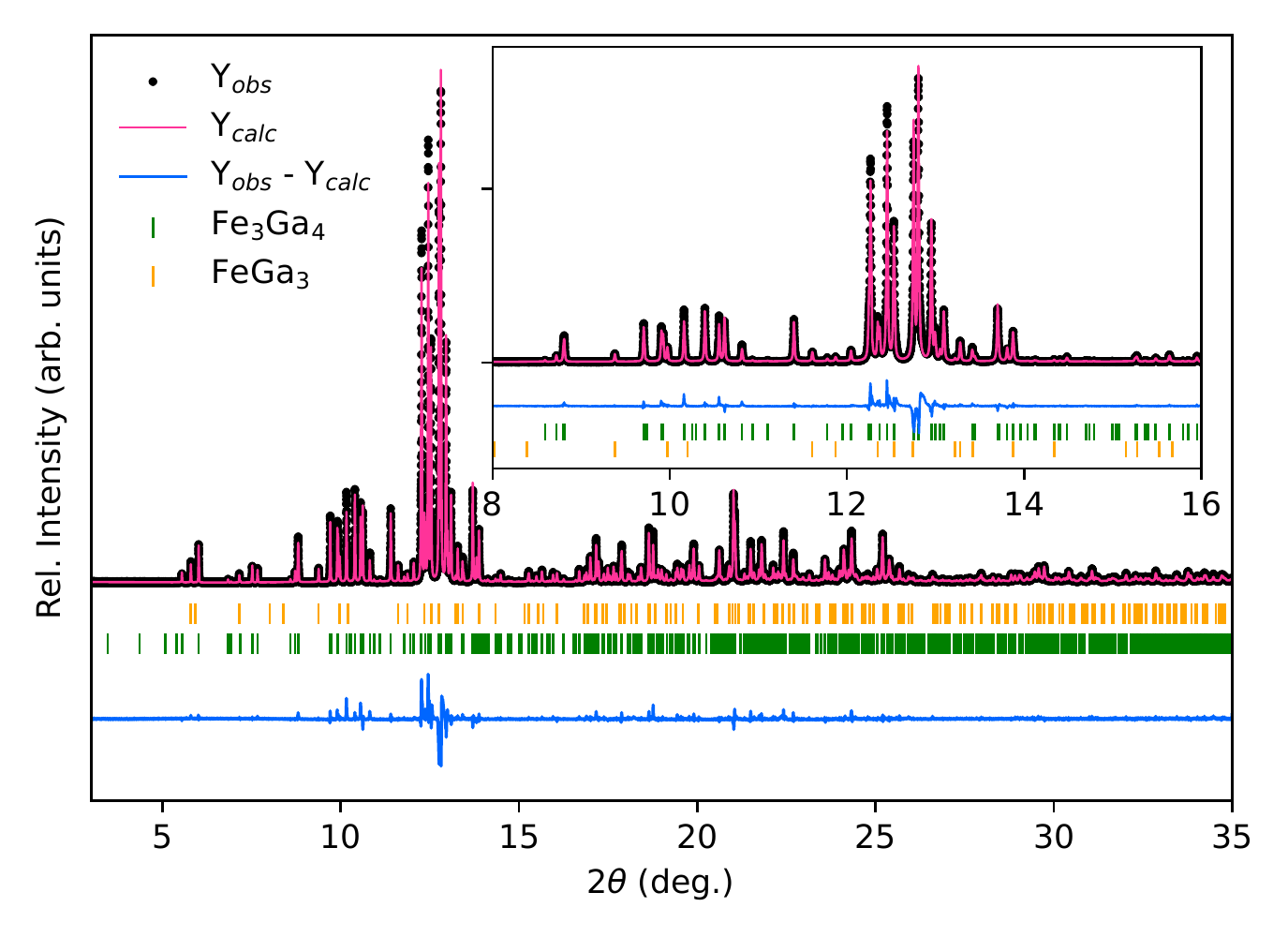}
    \caption{High-resolution synchrotron powder XRD pattern of 10\% V-doped Fe$_{3}$Ga$_{4}$ (Fe$_{2.7}$V$_{0.3}$Ga$_{4}$) collected at room temperature ($\sim$~295~K). Tick marks representing the corresponding binary phases are shown below the calculated, observed, and difference curves from Rietveld analysis. Fit statistics are summarized in the figure.} 
    \label{fig:xrd}
\end{figure*}

\begin{table*}[]
\caption{Crystallographic and composition analysis of arc-melted and annealed V-doped (Fe$_{1-x}$V$_{x}$)$_{3}$Ga$_{4}$ from Rietveld refinement of synchrotron powder XRD data. The first six columns display the crystallographic parameters for Fe$_{3}$Ga$_{4}$ for each V-doping concentration. The final two columns shown the evolution of the magnetic transitions in Fe$_{3}$Ga$_{4}$ with V-doping. All errors shown in parentheses for corresponding quantities. All blank cells indicate the transition was not observed in the measured temperature regime.}
\label{tab:my-table}
\begin{tabular}{|c|c|c|c|c|c|c|c|c|c|c|}
\hline 
\% V    & a (\AA)              & b (\AA)               & c (\AA)                & $\beta$ ($^{\circ}$)& Volume  (\AA$^{3}$)  & $x$ from EDS (\%). & \% FeGa$_{3}$ & T$_{1}$ (K)  & T$_{2}$ (K)  \\[2pt] \hline \hline
0 \%    & 10.1023(1)           & 7.6692(2)             & 7.8750(5)              & 106.29(1)           & 585.66(1)            & -              & 3.11(2)       &  68          &  360       \\[2pt] \hline
2.5 \%  & 10.1061(1)           & 7.6701(1)             & 7.8812(1)              & 106.26(1)           & 586.47(2)            & 2.4(1)         & 2.99(2)       &  42          &  343           \\[2pt] \hline
5 \%    & 10.1122(2)           & 7.6724(1)             & 7.8887(2)              & 106.23(2)           & 587.67(2)            &  4.7(4)        & 0.65(1)       &  57          &  282             \\[2pt] \hline
7.5 \%  & 10.1174(2)           & 7.6764(3)             & 7.8932(1)              & 106.19(2)           & 588.69(1)            & 7(1)           & 3.26(2)       &  48          &  252            \\[2pt] \hline
10 \%   & 10.1235(1)           & 7.6817(2)             & 7.8984(1)              & 106.16(2)           & 589.93(1)            &  11(1)         & 4.64(2)       &  13          &  182              \\[2pt] \hline
15 \%   & 10.1343(2)           & 7.6947(2)             & 7.9074(3)              & 106.11(1)           & 592.40(2)            &  14(1)         & 2.79(3)       &  -           &  75             \\[2pt] \hline
20 \%   & 10.1424(3)           & 7.7064(1)             & 7.9141(3)              & 106.08(3)           & 594.35(1)            &  20(2)         & 1.33(1)       &  -           &  70             \\[2pt] \hline
\end{tabular}
\end{table*}


Polycrystalline samples of V-doped (Fe$_{1-x}$V$_{x}$)$_{3}$Ga$_{4}$ were synthesized via arc-melting the elements in stoichiometric ratios with concentrations up to x = 0.20, using an Edmund Buehler MAM-1 under ultra-high purity Ar atmosphere. Elemental Fe granules (Alfa Aesar 99.98\%), Ga solid (Alfa Aesar 99.99\%) and V plates (Alfa Aesar, 99.7\%) were used. The arc-melted ingots were loaded into quartz tubes and sealed under moderate vacuum ($<10^{-3}$~Torr). These samples were annealed at 1000~$^\circ$C for 48~hr and allowed to cool to room temperature naturally. 

High-resolution synchrotron X-ray diffraction was performed on powders of ground as-recovered ingots at Beamline 11-BM at the Advanced Photon Source at Argonne National Lab through the mail-in program. Ground powders were packed in 0.4 mm Kapton capillary tubes under atmospheric conditions. Diffraction data was collected between 0.5 and 45 degrees with a step size of 0.0001$^{\circ}$, using a constant wavelength of 0.4581750 \AA~at 300~K. Rietveld refinements and data analysis was performed using the GSAS software suite.\cite{GSAS} Energy dispersive spectroscopy (EDS) spectra were collected on a Supra-25 SEM and Bruker EDS system at a beam energy of 20~kV and composition was determined as an average of different sites collected from each sample.

Magnetization of V-doped Fe$_{3}$Ga$_{4}$ was measured as a function of applied field M(H) and temperature M(T) using a Quantum Design SQUID MPMS system. For M(T), samples were first saturated in a field of 0.10~T, then the field was removed and the sample was cooled to 5~K. A field of 0.010~T was then applied and the magnetic moment was measured while heating to 400~K to obtain the zero-field-cooled (ZFC) curve. Isothermal M(H) measurements were performed to obtain magnetic hysteresis loops between $\pm$5~T at 10, 100, 200, 300 and 390~K for all doped samples.


\begin{figure}[]
    \centering
    \includegraphics[]{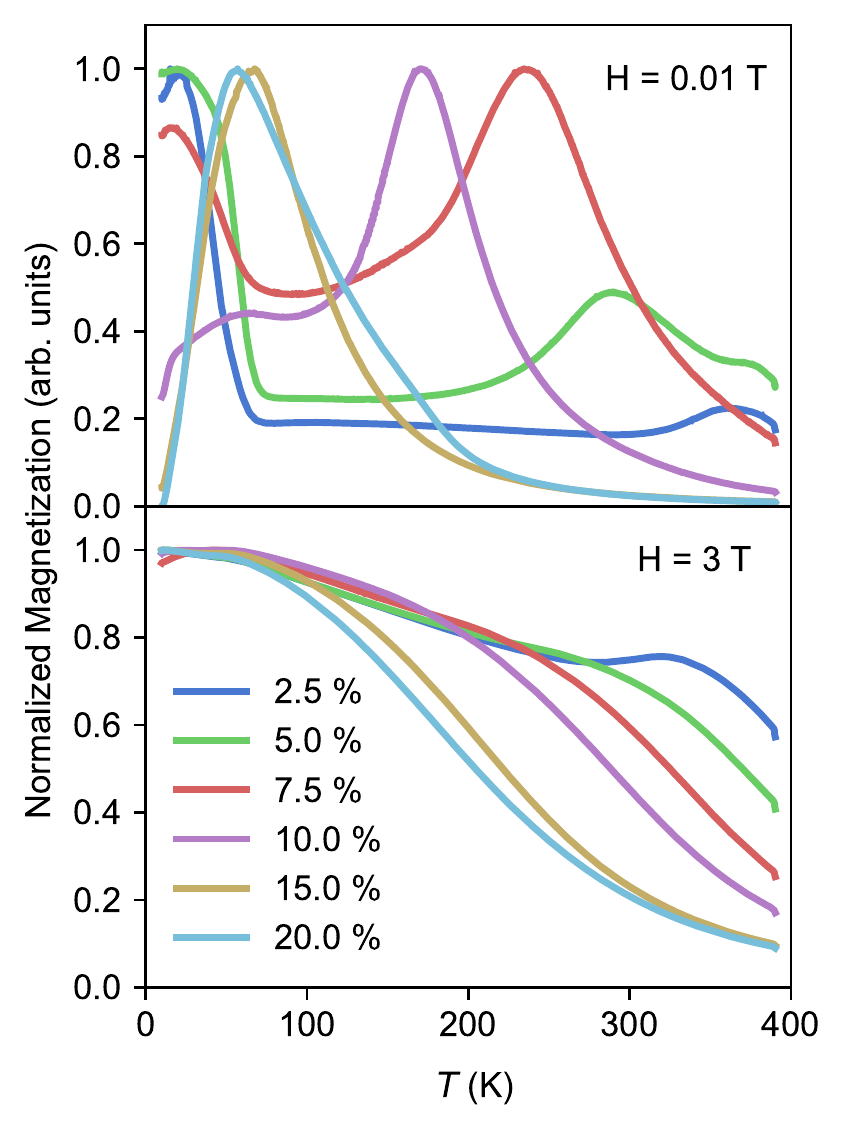}
    \caption{Normalized magnetization for all V-doped Fe$_{3}$Ga$_{4}$ samples as a function of temperature measured between 10 K and 400~K, under an external applied field of (a) 0.01~T and (b) 3~T. Only the ZFC curve for each measurement is shown for clarity. See supplementary materials for more information.} 
    \label{fig:XvT}
\end{figure}

\begin{figure*}[]
    \centering
    \includegraphics[]{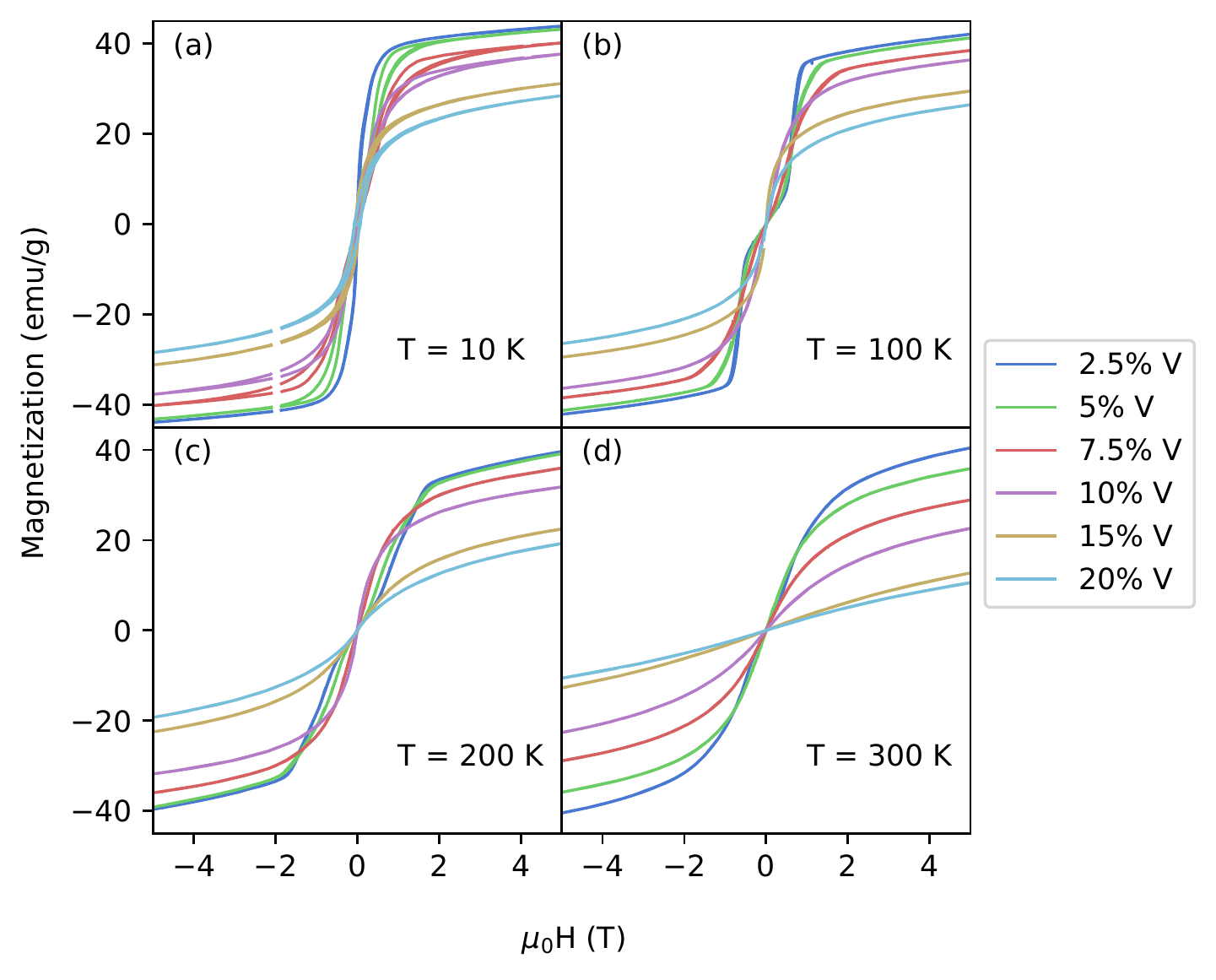}
    \caption{Isothermal magnetization for all V-doped samples as a function of applied field between -5 and 5 T, at four different temperatures: (a) 10 K, (b) 100 K, (c) 200 K, and (d) 300 K.} 
    \label{fig:MvH}
\end{figure*}

It has been shown that the magnetic order in Fe$_{3}$Ga$_{4}$ is sensitive to structural and electronic changes induced by annealing, pressure, and substitution.\cite{Wilfong2021, Al1995, Al2000, Duijn, Kobeissi, Al1998} Thus, synchrotron investigation of the structural changes induced by V-doping was undertaken. Results from Rietveld analysis of high-resolution synchrotron X-ray diffraction and composition analysis from EDS for all samples is summarized in Table \ref{tab:my-table}, and a representative X-ray diffraction pattern and Rietveld refinement is shown in Figure \ref{fig:xrd} for the 10\% V-doped sample. As seen in previous work focusing on polycrystalline synthesis, the Fe$_3$Ga$_4$ compound forms in the monoclinic C2/m structure and FeGa$_3$ is a secondary phase that can be minimized by annealing at 1000$^\circ$ C.\cite{Wilfong2021} FeGa$_3$ is a tetragonal structure with no associated magnetic behavior, leading to the magnetic properties due solely to the Fe$_3$Ga$_4$ compound.\cite{Samatham,Al1995} Table \ref{tab:my-table} illustrates the evolution of the structural parameters with respect to V-doping. As expected, based on the metallic radii of V and Fe, as V is substituted for Fe in Fe$_{3}$Ga$_{4}$, the lattice expands and overall the volume of the unit cell increases linearly following Vegard's Law. In addition, the diffraction confirms that V was incorporated into the Fe$_{3}$Ga$_{4}$ lattice. However, attempts to refine Fe site occupancy and distinguish between Fe and V were unsuccessful with the current data, but this remains a subject of further investigation using other techniques such as neutron powder diffraction. It was also confirmed that there were no V-segregate phases. Therefore, it was established that the V-doping leads to chemical strain, equivalent to negative pressure, and will subsequently affect the magnetic order.

Magnetization measurements were carried out to characterize how V-doping effects the AFM/SDW and FM phases, especially the transition temperatures separating these phases. The evolution of the magnetic order is seen in the normalized M(T) and the M(H) behavior presented in Figs. \ref{fig:XvT} and \ref{fig:MvH}. Figure \ref{fig:XvT}(a) shows M(T) for V-doped samples at a low applied field, which reveals changes in both the low-temperature FM-AFM transition T$_{1}$ and the high-temperature AFM-FM transition T$_{2}$. Qualitatively, it is seen that the temperature region of the AFM/SDW phase occurring between T$_{1}$ and T$_{2}$ narrows with increased V-doping. The quantitative results are summarized in Table \ref{tab:my-table}, where the changes in T$_{1}$ and T$_{2}$ are determined by the inflection points of the derivative of M(T), and shown in more detail in supplementary material Information. Importantly, the narrowing of the AFM/SDW phase with increasing V-doping, seen in Fig. \ref{fig:XvT}(a), is similar to the \textit{magnetic-field-induced} narrowing in undoped Fe$_{3}$Ga$_{4}$. The metamagnetic behavior - signified by magnetic-field-dependent transitions - is found in both undoped and V-doped Fe$_{3}$Ga$_{4}$. Also in both cases, the AFM/SDW phase narrows with increasing field. Figure \ref{fig:XvT}(b) shows that the AFM/SDW phase can even be eliminated in a moderate field of 2 to 3 T. 

Next, we focus on the similarity of the narrowing caused by V-doping with that caused by a magnetic field in undoped Fe$_{3}$Ga$_{4}$.\cite{WilfongSC,Wilfong2021} The main difference is seen in how the two bracketing transitions, T$_{1}$ and T$_{2}$, evolve. With increasing field in undoped Fe$_{3}$Ga$_{4}$ the narrowing was dominated by a 200~K \textit{increase} in T$_{1}$; whereas for V-doping the narrowing is dominated by a 200~K \textit{decrease} in T$_{2}$, shown in Fig. \ref{fig:XvT}(a). For V-doping, T$_{2}$ decreases from 360 to 75~K at 15\% V-doping (see Table \ref{tab:my-table}). But opposite to the undoped case where T$_{1}$ shows a large increase, V-doping decreases T$_{1}$ mildly from 68 to 13~K at 10\% doping. Note that the transition does appear to broaden with increased V content, thus impacting the effective T$_{1}$. For the 10\% V-doped sample, T$_{1}$ is barely observable, and for higher V concentrations the low-temperature FM phase is no longer detected.

Transitions of the AFM/SDW phase are observed in isothermal M(H) measurements, shown in Fig. \ref{fig:MvH}, which appear as kinks in the M(H). At 100~K, the low-doped samples are within the AFM/SDW phase at zero field, but at higher fields Fig. \ref{fig:MvH}(b) shows a clear metamagnetic transition for the 2.5 and 5\% V-doped samples. For the 15 and 20\% V-doped samples, the metamagnetic behavior is completely suppressed. A similar evolution is present in the 200~K data in Fig. \ref{fig:MvH}(c), but is not as clear at this higher temperature. The 15 and 20\% samples show no metamagnetic behavior, similar to that for high fields in undoped Fe$_{3}$Ga$_{4}$. Overall, Fig. \ref{fig:MvH} shows that for all temperatures the moment near saturation decreases with increasing V-doping, which is attributed to decreasing Fe content and the Fe-Fe coupling. Further details of the M(H) behavior can be found in the supplementary material. 

In conclusion, we have studied the effects of vanadium doping on the structural and magnetic properties of polycrystalline Fe$_{3}$Ga$_{4}$. Through structural work, it was found that V-doping up to 20\% results in no magnetic impurity phases and a systematic increase in unit cell volume upon increasing V-doping. In addition, elemental analysis confirmed V substitution into the Fe$_{3}$Ga$_{4}$ lattice. Temperature-dependent magnetization measurements showed significant changes in the T$_{1}$ and T$_{2}$ transition temperatures that bracket the helical AFM/SDW phase. T$_{1}$ was only slightly affected by V-doping up to 7.5\% nominal doping. In contrast, the T$_{2}$ transition was a strong function of V-doping, decreasing from 360~K to $\sim$~150~K at 10\% doping. For both undoped and V-doped Fe$_{3}$Ga$_{4}$, an applied magnetic field of only 2~T transforms the AFM/SDW phase into a ferromagnetically-polarized phase. However, it was found that above 10\% V-doping the ground state is FM at all temperatures. The changes in magnetic behavior upon V-doping in Fe$_3$Ga$_4$ are likely due to the changes in both localization of electron and magnetic interaction as well as the tuning of the electronic properties and electronic bands at the Fermi level which are both theoretically predicted to be integral to the complex magnetic order in Fe$_3$Ga$_4$. In summary, several points are noteworthy about the magnetic and structural properties of the AFM/SDW phase resulting from V-doping: (1) V-doping and magnetic field are similar in that they both narrow the AFM/SDW phase; and (2) the narrowing effect of V-doping has the effect of \textit{negative} pressure, as V-doping causes the lattice to expand, whereas \textit{positive} applied pressure is known to narrow the AFM/SDW phase. This work demonstrates that doping is an effective way to tune the magnetic phase diagram of Fe$_{3}$Ga$_{4}$ for consideration and optimization in devices, as well as to further characterize the magnetic phases of Fe$_{3}$Ga$_{4}$.

\begin{acknowledgements}
Research at the United States Naval Academy was supported by the NSF DMR-EPM 1904446 and ONR 1400844839. Work at Northeastern University was partially supported by the National Science Foundation grant DMR-1905662 and the Air Force Office of Scientific Research award FA9550-20-1-0247 (A.F. and D.H.). Work at VCU was partially funded by National Science Foundation, Award Number: 1726617.
\end{acknowledgements}

\label{References}
\bibliographystyle{apsrev4-1}
\bibliography{bib}
\end{document}